\documentclass[a4paper,12pt]{article}
\pdfoutput=1
\usepackage{graphicx,rotating,amssymb}

\ifx\pdfoutput\undefined
\usepackage[dvips,bookmarks]{hyperref}	
\else
\usepackage{hyperref}	
\fi
\hypersetup{colorlinks,bookmarksopen,bookmarksnumbered,
linkcolor=blus,pdfstartview=FitH,urlcolor=rossos,citecolor=verdes}
\def\hhref#1{\href{http://arxiv.org/abs/#1}{#1}} 

\newcommand{\fig}[1]{~\ref{fig:#1}}
\usepackage{multicol}
\usepackage{color}
\definecolor{rosso}{cmyk}{0,1,1,0.4}
\definecolor{rossos}{cmyk}{0,1,1,0.55}
\definecolor{rossoc}{cmyk}{0,1,1,0.2}
\definecolor{blu}{cmyk}{1,1,0,0.3}
\definecolor{blus}{cmyk}{1,1,0,0.6}
\definecolor{bluc}{cmyk}{1,1,0,0.1}
\definecolor{verde}{cmyk}{0.92,0,0.59,0.25}
\definecolor{verdec}{cmyk}{0.92,0,0.59,0.15}
\definecolor{verdes}{cmyk}{0.92,0,0.59,0.4}
\newcommand{\nubarnu}{\raisebox{1ex}{\hbox{\tiny(}}\overline\nu\raisebox{1ex}{\hbox{\tiny)}}\hspace{-0.5ex}}

\newcommand{\eq}[1]{~{\rm (\ref{eq:#1})}}

\newcommand{\GeV}{\,{\rm GeV}}
\newcommand{\TeV}{\,{\rm TeV}}
\newcommand{\cm}{\,{\rm cm}}

\def\circa#1{\,\raise.3ex\hbox{$#1$\kern-.75em\lower1ex\hbox{$\sim$}}\,}

\newcommand{\beq}{\begin{equation}}
\newcommand{\eeq}{\end{equation}}

\newcommand{\bea}{\begin{eqnarray}}
\newcommand{\eea}{\end{eqnarray}}

\font\tenrsfs=rsfs10 at 12pt
\font\sevenrsfs=rsfs7
\font\fiversfs=rsfs5
\newfam\rsfsfam
\textfont\rsfsfam=\tenrsfs
\scriptfont\rsfsfam=\sevenrsfs
\scriptscriptfont\rsfsfam=\fiversfs
\def\mathscr#1{{\fam\rsfsfam\relax#1}}

\def\circa#1{\,\raise.3ex\hbox{$#1$\kern-.75em\lower1ex\hbox{$\sim$}}\,}
\makeatletter

\def\art{\@ifnextchar[{\eart}{\oart}}
\def\eart[#1]#2#3#4#5#6{{\rm #2}, {#3 #4} {\rm (#6) #5} [arXiv:{\hhref{#1}}]}
\def\hepart[#1]#2{{\rm #2, arXiv:\hhref{#1}}}
\newcounter{alphaequation}[equation]
\def\thealphaequation{\theequation\hbox to
0.6em{\hfil\alph{alphaequation}\hfil}}
\def\eqnsystem#1{
\def\@eqnnum{{\rm (\thealphaequation)}}
\def\@@eqncr{\let\@tempa\relax \ifcase\@eqcnt \def\@tempa{& & &} \or
 \def\@tempa{& &}\or \def\@tempa{&}\fi\@tempa
 \if@eqnsw\@eqnnum\refstepcounter{alphaequation}\fi
\global\@eqnswtrue\global\@eqcnt=0\cr}
\refstepcounter{equation} \let\@currentlabel\theequation \def\@tempb{#1}
\ifx\@tempb\empty\else\label{#1}\fi
\refstepcounter{alphaequation}
\let\@currentlabel\thealphaequation
\global\@eqnswtrue\global\@eqcnt=0 \tabskip\@centering\let\\=\@eqncr
$$\halign to \displaywidth\bgroup \@eqnsel\hskip\@centering
$\displaystyle\tabskip\z@{##}$&\global\@eqcnt\@ne
\hskip2\arraycolsep\hfil${##}$\hfil& \global\@eqcnt\tw@\hskip2\arraycolsep
$\displaystyle\tabskip\z@{##}$\hfil
\tabskip\@centering&\llap{##}\tabskip\z@\cr}
\def\endeqnsystem{\@@eqncr\egroup$$\global\@ignoretrue} \makeatother

\newcommand{\SU}{\,{\rm SU}}

\begin{document}

\begin{center}
IFUP-TH/2008-27\hfill SACLAY--T08/139

\bigskip

{\LARGE\bf\color{magenta}
Model-independent implications  \\
of the $e^\pm$, $\bar p$ cosmic ray spectra
 \\
on properties of Dark Matter \\[3mm] \color{black}
(updated including AMS 2013 data)
}\\

\medskip
\bigskip\color{black}\vspace{0.6cm}
{
{\large\bf M. Cirelli$^a$, M. Kadastik$^b$, M. Raidal$^b$, A. Strumia$^c$}
}
\\[7mm]
{\it $^a$ Institut de Physique Th\'eorique, CEA-Saclay and CNRS, France\footnote{CEA, DSM, Institut de Physique Th\`eorique, IPhT, CNRS, MPPU, URA2306, Saclay, F-91191 Gif-sur-Yvette, France}} \\
{\it $^b$NICPB, Ravala 10, 10143 Tallinn, Estonia} \\
{\it $^c$Dipartimento di Fisica dell'Universit{\`a} di Pisa and INFN, Italia}

\bigskip\bigskip\bigskip

{\large
\centerline{\large\bf Abstract}

\begin{quote}
\small
Taking into account spins, we classify all two-body non-relativistic Dark Matter 
annihilation channels to the allowed polarization states of Standard Model particles, computing the energy spectra of the stable final-state particles relevant for indirect DM  detection. 
We study the DM masses, annihilation channels and cross sections that can 
reproduce the PAMELA  indications of an $e^+$ excess consistently  
with the PAMELA $\bar p$ data and the ATIC/PPB-BETS $e^++e^-$ data. 
From the PAMELA data alone, two solutions emerge: 
$(i)$ either the DM particles that annihilate
into $W,Z,h$ must be heavier than about 10 TeV or  $(ii)$ the  DM must annihilate only into leptons.
Thus in both cases a DM particle compatible with the PAMELA excess seems to have quite unexpected properties.  The solution $(ii)$ implies a  peak in the  $e^++e^-$ energy spectrum, 
 which, indeed, seems to appear in the ATIC/PPB-BETS data around 700 GeV. 
 If upcoming data from ATIC-4 and GLAST confirm this feature, this would point to a ${\cal O}(1)$~TeV DM annihilating only into leptons. Otherwise
 the solution $(i)$ would be favored.
We comment on the implications of these results for DM models, direct DM detection and colliders as well as on the possibility of an astrophysical origin of the excess.

\end{quote}}

\end{center}

\pagebreak

\section{Introduction and highlights}
Cosmological observations imply that about $80\%$ of the mass of the Universe
is some unknown form of cold 
Dark Matter (DM)~\cite{Komatsu:2008hk}. Presently the origin and 
nature of the DM particles, their mass, spin, couplings and other properties remain 
completely unknown.  Among many possible cold DM candidates the most popular 
ones are the stable weakly interacting massive particles (WIMPs) which occur in many
extensions of the Standard Model (SM), most notably in supersymmetry.  If the DM WIMPs
are thermal relics,  their annihilation cross section must be 
$\sigma v \sim 3\cdot 10^{-26}\cm^3/{\rm sec}$,
which indeed is typical of a weakly interacting TeV-scale particle.
To discover the DM particles, experiments have searched for direct production 
of WIMPs in collider experiments, their scattering off the nuclei in terrestrial detectors 
as well as {\it indirect signals} of particles generated by the DM annihilation  
in the galactic halo~\cite{pioneers}. 

The recently reported  results by the PAMELA experiment~\cite{PAMELA}
have the opportunity, if confirmed, of establishing a breakthrough in the cosmic antimatter searches.  
The  PAMELA data show 
$(i)$ a steep increase in the energy spectrum of
the positron fraction, $e^+/(e^++e^-),$  in cosmic rays above  10~GeV~\cite{PAMELA}, compatibly with previous less certain hints from HEAT~\cite{Barwick:1997ig} and AMS-01~\cite{Aguilar:2007yf};
$(ii)$ no excess in the $\bar p/p$  energy spectrum~\cite{PAMELApbar} compared with the predicted background;
$(iii)$ at low energy, $E_{e^+}< 10$~GeV,  the positron flux is presently suppressed by the  
solar magnetic polarity state $A^-$~\cite{clem}. 
The positron excess observed by PAMELA is so big that
it can produce observable spectral features in the total $e^++e^-$ flux.
Indeed, most recently the PPB-BETS balloon experiment reported 
an excess in the $e^++e^-$ energy spectrum between 500-800~GeV \cite{Torii:2008xu},
confirming the similar earlier claim by the ATIC-2 balloon experiment \cite{ATIC-2}. 
These data-sets have not received much attention yet as probes of DM indirect detection, but, 
as we will see, they have the potential of being relevant to the issue.
Of course one should be careful: the Monte Carlo simulations that such experiments need
to tag $e^\pm$ and infer their energy have been tested only up to LEP energies;
the excess is based on just a few data-points that are not cleanly consistent between ATIC-2 and the smaller PPB-BETS;
emulsion chambers (EC) balloon experiments~\cite{EC} do not show evidence for an excess, although they have larger uncertainties.
The excess in both experiments seems to shows a cut-off at energies just below 1~TeV. 
Those results will soon be tested by forthcoming ATIC-4 \cite{ATIC-4} 
and GLAST/Fermi data~\cite{GLAST}, that have bigger calorimeters able of containing a electromagnetic shower.
All these excesses might be of course due to some astrophysical source, such as a nearby pulsar~\cite{pulsar}.
However, they also might provide the first clear evidence for the DM annihilations into SM particles.

In this work we explore whether galactic DM annihilations can account for the observed 
excesses in the data.
We perform a model independent analysis of the PAMELA positron 
and antiproton data with and without the electron plus positron data obtained from the balloon
experiments.
Our aim is to study if and how one can get model independent 
information on the DM mass, spin,
interactions to the SM particles and on the DM annihilation cross section from the 
experimental data.
Taking into account restrictions that must be obeyed by 
non-relativistic DM annihilations, we identify the parameters that can be probed by the indirect signals.
We study the non-relativistic DM annihilation cross sections into the set of all possible SM particle
final states, DM DM $\to$ SM SM, where $${\rm SM}=\{e, \mu_L, \mu_R, \tau_L, \tau_R, W_L,  W_T, Z_L, Z_T, h, q, b, t\},$$
taking into account the allowed polarizations
($T$ransverse, $L$ongitudinal, $L$eft, $R$ight).
Using Monte Carlo tools, partly written by us, we try to keep 
the polarizations/helicities of the DM annihilation products and 
 compute the energy spectra of the
final state $e^\pm,p^\pm,\gamma, \nubarnu_{e,\mu,\tau}$
coming from their decays.
Indirect signals of any DM model can be obtained
by combining  our decay spectra according to the branching ratios predicted by the model.
The experimental results can be analyzed in terms of these phenomenological parameters,
without committing to some specific theoretical scenario.

\begin{figure}[p]\begin{center}
$$\includegraphics[width=\textwidth]{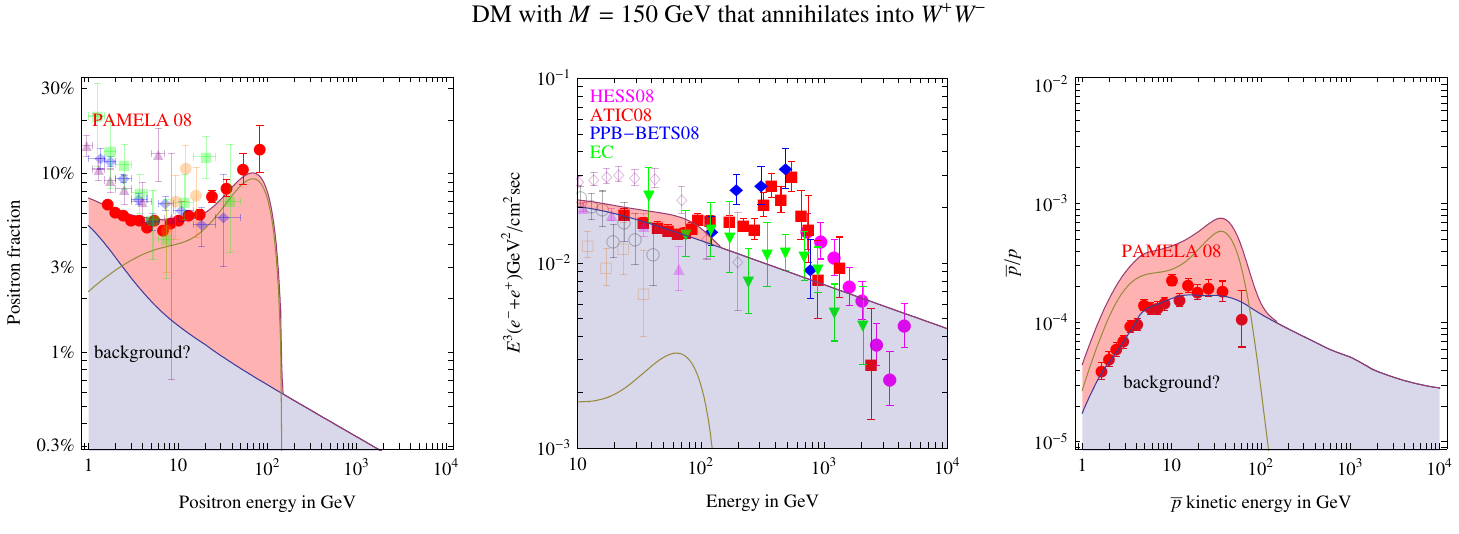}$$
$$\includegraphics[width=\textwidth]{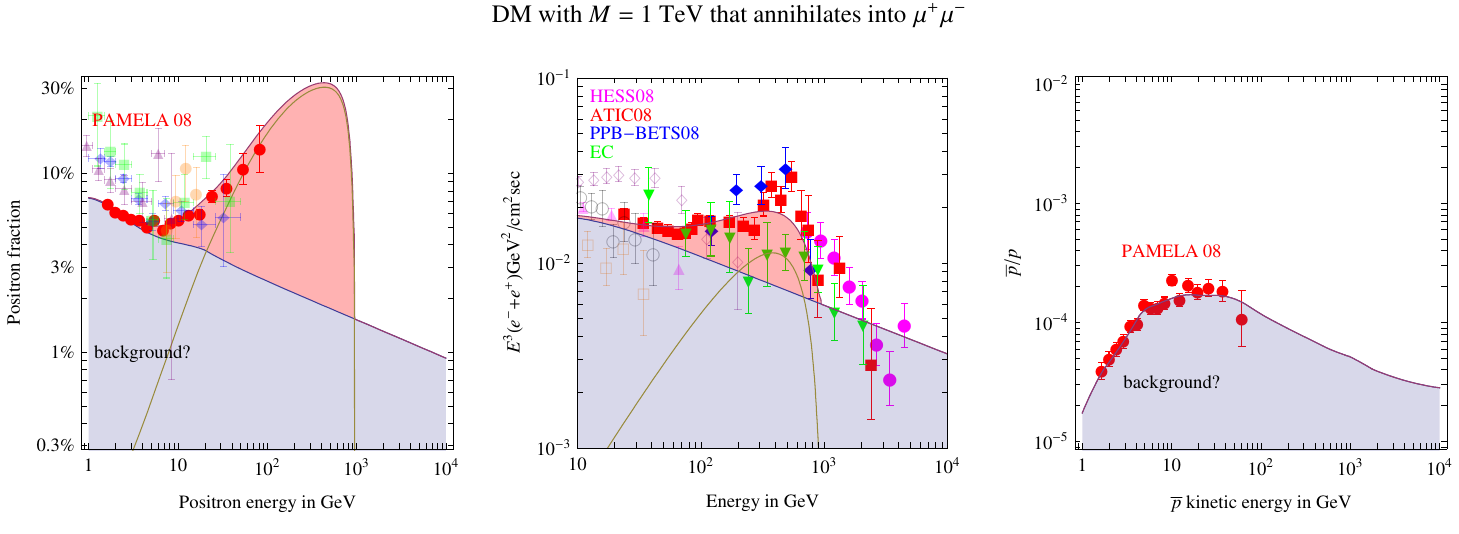}$$
$$\includegraphics[width=\textwidth]{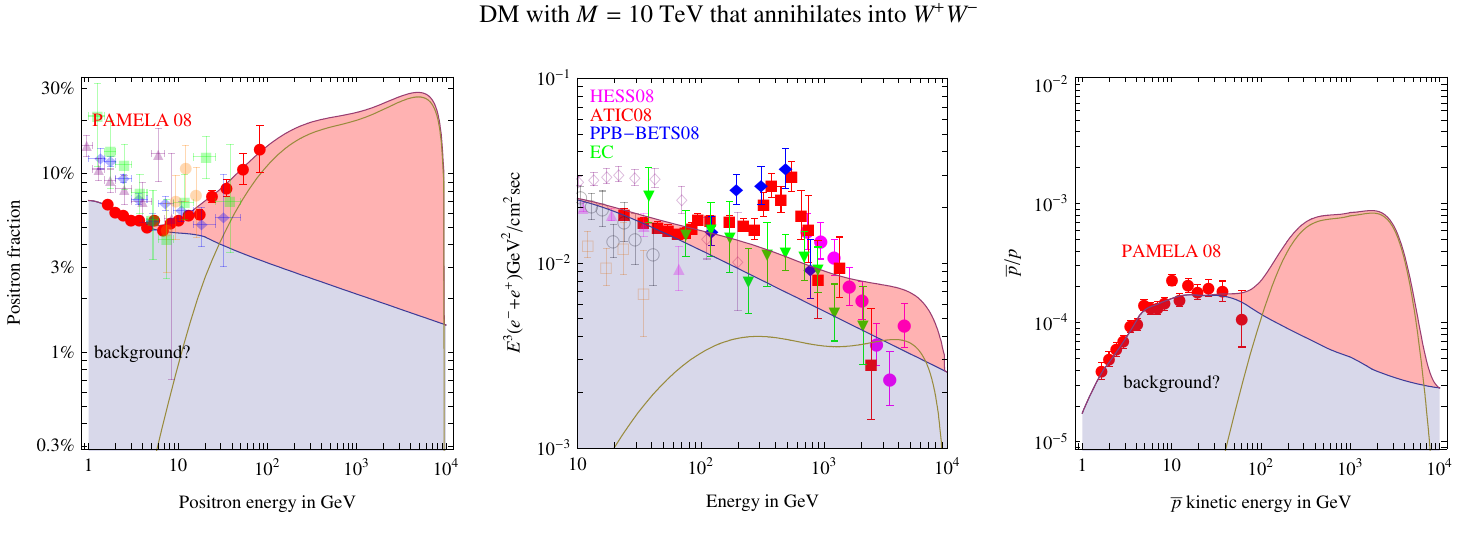}$$
\caption{\em Three examples of fits of $e^+$ (left), $e^++e^-$ (center), $\bar{p}$ (right) data,
for $M=150\GeV$ (upper row, excluded by $\bar{p}$), $M=1\TeV$ (middle row, favored by data),
$M=10\TeV$ (lower row, disfavored by the current  $e^++e^-$ excess).
Galactic DM profiles and propagation models are varied to provide the best fit. See Sec.~\ref{sec:propagation} for the discussion on the treatment of the uncertain astrophysical background. 
\label{fig:samples}}
\end{center}
\end{figure}

Taking into account different models for the DM distribution 
and for the electron and proton propagation in the Galaxy, 
we perform fits of the predicted $e^+/(e^++e^-),$ $\bar{p}/p$ and $(e^++e^-)$
energy spectra  to the PAMELA and balloon data. 
We allow for the possibility of enhanced annihilation cross sections and/or for the 
high density substructures in the DM halo, assuming an energy-independent boost factor $B.$ 
 For simplicity, we assume a common boost factor for the $e^\pm$ and for the $p^\pm$ and we take it as energy-independent. These are widely used assumptions, but they are not the only possibility~\cite{Lavalle}. One can easily infer what happens if these assumptions are relaxed. 
We thus study which DM masses and annihilation channels can best explain the experimental measurements.

\medskip

To illustrate our results, as well as to present the PAMELA and balloon
$e^+/(e^++e^-),$  $(e^++e^-)$ and $\bar{p}/p$ data,
we show in Fig.~\ref{fig:samples} three examples for the DM masses $M =150$~GeV,
1~TeV and 10~TeV that annihilate into $W^+W^-$ and $\mu^+\mu^-,$ as indicated 
in the figures.
The first column presents the predictions for the positron fraction, the second for the total
electron and positron flux and the third for the antiproton over proton flux.

Many DM annihilation channels into SM primary channels can give a reasonably good fit to the PAMELA $e^+/(e^++e^-)$ excess alone, at least for some value of the DM mass picked in a wide range 60 GeV $\to$ tens of TeV
(first column of Fig.~\ref{fig:samples}; see Fig.~\ref{fig:fite} for the full analysis). 
More precisely, DM annihilations into SM leptons can well fit for almost any mass in this range, while annihilations into quarks and Higgs bosons are disfavored at low masses because of the soft positron spectrum. 
Annihilations into gauge bosons work for light and heavy masses. 

The inclusion of the $\bar{p}$ data changes the issue.  
Because the DM annihilations to gauge bosons, Higgs bosons 
or quarks ($W,Z,h,q,b,t$) produce antiprotons, 
under our assumptions the PAMELA $\bar{p}/p$ data excludes 
light DM with sizable annihilation fraction into these channels (Fig.~\ref{fig:samples}, the first row). 
However, for $M \circa{>} 10$~TeV the antiproton 
spectrum becomes again consistent with present data, and the $W^+W^-$ seem to give the relatively best fit (Fig.~\ref{fig:samples}, the third row). 

We find therefore that the PAMELA data  single out two sets of solutions which
can satisfy the observations compatibly with our standard astrophysical assumptions: 
(i) a heavy ($\circa{>}$ 10 TeV) DM that annihilates dominantly into $W^+W^-$; (ii) a DM that annihilates into SM leptons, with no strong preference for any DM mass.

\medskip

The inclusion of the balloon data qualitatively changes the sensitivity to the DM mass
because the PAMELA excess predicts observable spectral features in the total flux 
(Fig.~\ref{fig:samples}, the second column).
It is striking  that the same signal of the DM annihilation to  leptonic
channels which can explain the PAMELA data  is also able to produce the apparent 
$(e^++e^-)$ excess in the PPB-BETS and ATIC data. Because the balloon
data shows a sharp cut-off in the excess just below 1 TeV, the DM mass
should be close to $1$~TeV, and all other but leptonic DM annihilation 
channels would be strongly disfavored or excluded. On the other hand, if the $e^++e^-$ excess
will not be confirmed, the $e^++e^-$ data will strongly constrain or exclude the 
DM annihilations to leptons.  Should this be the result of  the upcoming ATIC-4 (or GLAST) experiment, 
the astrophysically favorite candidate for DM becomes a heavy $\sim$10 TeV WIMP
annihilating into SM gauge bosons.  

We comment on the implications of PAMELA and balloon data 
for supersymmetric DM candidates~\cite{Ellis:1983ew},
as well as for some theoretically more exotic  alternatives~\cite{Bertone:2004pz,Hooper:2007qk}.
We explore in a qualitative way the model-independent expectations for direct DM 
searches and for the 
production and detection  of the DM particles at LHC and future lepton colliders.  
A few authors have discussed implications of the PAMELA $e^+/(e^++e^-)$  
results~\cite{Bergstrom:2008gr, Cirelli:2008jk,Barger:2008su,Cholis:2008hb}.
While the analysis of the $e^+/(e^++e^-)$ data presented in~\cite{Barger:2008su,Cholis:2008hb}
is consistent with our more systematic treatment,
our final conclusions differ because we also include the important $\bar p$ and $e^++e^-$ data.


\bigskip

The paper is organized as follows. Section \ref{sec:cosmology} discusses cosmological 
constraints  on the DM annihilation cross section. In Section \ref{sec:theory}
we identify the DM annihilation channels and compute the spectra of SM final states. 
In Section \ref{sec:propagation} we discuss the particle propagation in the Galaxy.
In Sections \ref{sec:PAMELApositrons}, \ref{sec:PAMELAantiprotons} and \ref{sec:balloons} we perform fits to the PAMELA positron data, PAMELA positron plus
antiproton data, and to the $e^\pm$ PAMELA and balloon data, respectively.
We discuss the implications of our results for DM direct detection in Section \ref{sec:direct} and for
collider phenomenology in Section \ref{sec:collider}.
We conclude in Section \ref{sec:conclusions}.

\begin{figure}[t]
\begin{center}
$$\includegraphics[width=0.8\textwidth,height=9cm]{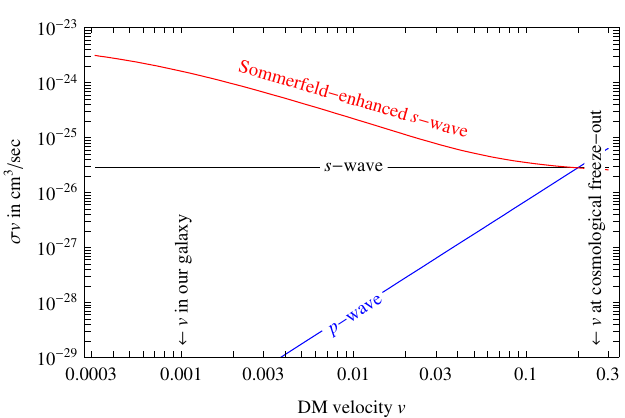}$$
\caption{\em Values of the DM annihilation cross sections suggested by the DM abundance.
\label{fig:sigmaDMDM}}
\end{center}
\end{figure}

\begin{figure}[t]
\begin{center}
$$\includegraphics[width=0.45\textwidth]{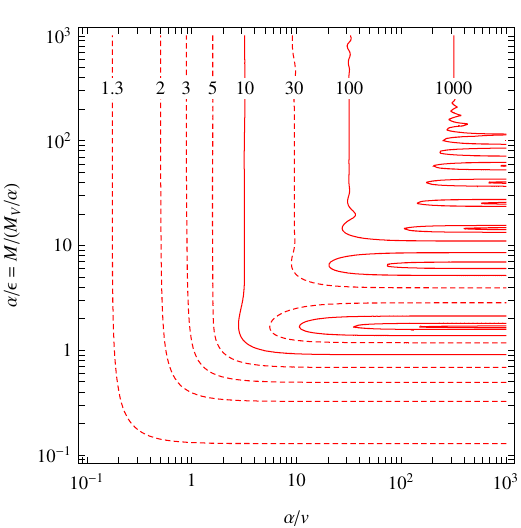}\qquad
\includegraphics[width=0.45\textwidth]{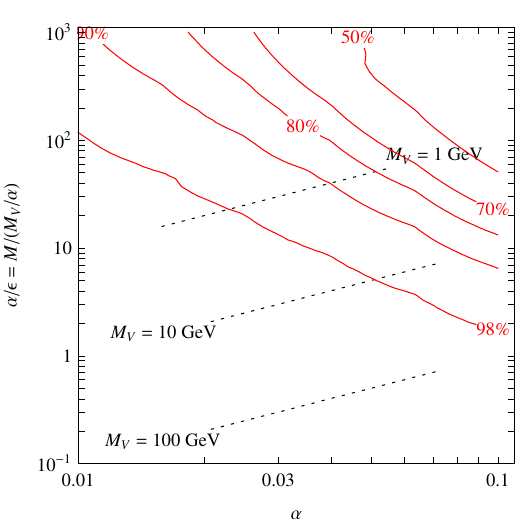}$$
\caption{\em Contour plots of the Sommerfeld enhancement for one abelian massive vector with mass $M_V$ and coupling $\alpha$ to DM (from~\cite{Sommerfeld2})
and the resulting Sommerfeld reduction in the cosmological thermal DM abundance
(we here assumed a DM mass $M=1\TeV$).
\label{fig:Som}}
\end{center}
\end{figure}

\section{Cosmology and the total DM DM cross section}
\label{sec:cosmology}
The indirect signal rate is proportional to the total DM annihilation cross section.
As well known, assuming that DM is a thermal relic, the measured DM cosmological abundance fixes
the total DM annihilation rate~\cite{Beltran:2008xg}.  Assuming that we can neglect co-annihilations with other new particles slightly heavier than DM, cosmology suggests
\beq \sigma v \approx 3\times10^{-26}\cm^3/{\rm sec}\qquad\hbox{around
freeze-out, i.e.\ $ v\sim 0.2$}. \eeq
Astrophysical indirect DM signals are proportional to $\sigma v$ at $ v\sim 10^{-3}$, which is
the typical DM velocity (with respect to $c$) in our Galaxy, as determined by the escape velocity independently of the DM mass $M$.
Fig.\fig{sigmaDMDM} illustrates the possible extrapolations of $\sigma v$ from $ v\sim0.2$ down to $ v\sim10^{-3}$:
\begin{itemize}
\item[a)] $\sigma v$ remains constant if, in the non-relativistic limit $ v\ll1$, the cross section is dominated by $s$-wave annihilations.
This leads to an indirect signal rate not much below the sensitivity of present experiments, that imply $\sigma v\circa{<}10^{-23}\cm^3/{\rm sec}$, depending on the branching ratios and on astrophysical issues.

\item[b)]
Alternatively, $p$-wave annihilations might dominate if $s$-wave annihilations are suppressed --- in view of $\sigma v \propto  v^2$  this leads to a negligibly small rate of indirect DM signals.
Case b) will therefore be  ignored in our analysis.

\item[c)]
The most favorable possibility  is that $s$-wave annihilations dominate and are enhanced,
in the non-relativistic limit, by the presence of Sommerfeld corrections, 
present in the SM if DM interacts with the $W,Z$ vectors with a coupling $g\sim g_2$
and is heavier than $M\circa{>} 4\pi M_W/g^2\sim 2\TeV$ ~\cite{Sommerfeld1,Sommerfeld2}.
In such a case $\sigma v$ roughly grows as $ v_{\rm max}/ v$ in the range  $v_{\rm min}< v <v_{\rm max}$ where
$ v_{\rm max} \approx g^2/4$ and $v_{\rm min}\approx M_W/M$~\cite{Sommerfeld2}\footnote{A smaller $v_{\rm min}$ is possible for specific values of $M$ that lead
to a zero-energy DM DM bound state; we estimate that DM annihilations in first structures~\cite{first} put the bound $v_{\rm min}\circa{>} 10^{-3}$,
compatibly with Minimal Dark Matter predictions, as shown in fig.~1 of~\cite{Sommerfeld2}.}
However details are model-dependent.
The red line in Fig.\fig{sigmaDMDM} illustrates the Sommerfeld enhancement obtained for $g=g_2$: 
astrophysical signals are enhanced by a few orders of magnitude with respect to case a).
\end{itemize}
 We here proceed focussing on $s$-wave annihilations, and we assume the standard case a)
 to  normalize our plots in Figs.\fig{espectra},\fig{pspectra},\fig{compspectra}.

However, as we will see, the PAMELA and ATIC/PPB-BETS anomalies suggest that $\sigma  v$ and/or
the electron boos factor $B_e$ is much larger than what is suggested by cosmology and/or astrophysics~\cite{Lavalle}.
Co-annihilations only allow to gain (or loose) ${\cal O}(1)$ factors.
Possible resonances (e.g.\ a particle with mass very close to $2M$)
may have a bigger effect but are strongly model dependent.
If future data will confirm that such a sizable enhancement is needed, 
and disregarding possible  resonances, 
one can pursue two possibilities:
i) DM is not a thermal relic, so that it can have a $\sigma v$ larger than what suggested by cosmology;
ii) DM is a thermal relic, and $\sigma v$ is Sommerfeld enhanced at low $ v$.
Within the SM this requires a DM mass heavier than a few TeV; in general one can
assume that DM interacts with some extra light vector or scalar with mass $M_V$ and coupling $\alpha$ to DM.
Fig.\fig{Som} illustrates that a sizable Sommerfeld correction at low velocity $v$
gives a relatively minor correction to the cosmological DM abundance.

%


\section{Possible DM annihilations}
\label{sec:theory}
We study how Lorentz and gauge-invariance restrict the possible non-relativistic DM annihilations.  
We assume that the primary annihilation products contain two SM particles, so that the possible cases
are
\beq \label{eq:list}
W^+ W^-,\quad ZZ,\quad Zh,\quad hh,\qquad e^+ e^-,\quad \mu^+\mu^-,\quad
\tau^+\tau^-,\qquad b\bar b,\quad t\bar t,\quad q\bar q,
\eeq
where $q$ denotes any light quark, $u,d,s,c$.
In view of the presumed neutrality of the DM particle we have not included final states containing photons.
Bounds from direct detection experiments~\cite{Ahmed:2008eu,Angle:2007uj,Bernabei:2008yi} suggest that the DM coupling to the $Z$ is much smaller than $g_2$, so that also the $ZZ$ primary channel should perhaps be excluded from the list in eq.\eq{list}.

\medskip

Next, we need to consider the allowed polarizations of the SM primary particles in eq.\eq{list},
because the energy spectra of final-state $e^\pm,p^\pm,\gamma,\nubarnu_{e,\mu,\tau}$ produced
by their decays depend on their polarization.
Since we assume that DM is an elementary weakly interacting particle, its spin is restricted to be 0 or 1/2 or 1.
In view of the spin-composition rules
\beq
 1 \otimes 1=1,\qquad
2\otimes 2 =1_A\oplus 3_S,\qquad
3\otimes 3 = 1_S\oplus 3_A\oplus 5_S,
\eeq
the spin of the two-body non-relativistic $s$-wave DM DM state can be 0, 1 or 2.
In the Majorana fermion case, the wave function must be asymmetric under exchange of the two DM particles,
so that only the $1_A$ two-body state is allowed.
In summary, for all DM spins the two-body DM DM state can have spin 0;
higher spin is possible only in some cases.

In order to be general and to avoid the lengthy expressions obtained from
 explicit evaluation of Feynman diagrams even in the simple DM models, 
 we use tools from effective relativistic quantum field theory.
Indeed, the non-relativistic nature of the DM DM two-body state is a relativistically invariant concept,
and its annihilation products are quantistic and relativistic, so that these concepts are needed.
In practice, all this amounts to say that the two body DM DM state is represented, depending on its spin, 
by a scalar field $\mathscr{D}$, or a vector $\mathscr{D}_\mu$, or $\mathscr{D}_{\mu\nu}$ if DM is a vector.
This latter case is somewhat special, as it needs one extra gauge invariance,
and is realized within Little Higgs models with $T$-parity as well as in models with extra dimensions.\footnote{
As the possible underlying motivations are not important for our purposes, we present a simplified
version of its  Lagrangian.
In the simplest case the gauge group is abelian, e.g.\ a KK excitation of hypercharge.
Taking into account this abelian factor of the SM, we introduce 
a gauge group $G={\rm U}(1)_1\otimes{\rm U}(1)_2$ and
three  Weyl fermions $\psi$, $\psi_1$ and $\psi_2$  with charges
$(1,1)$, $(-2,0)$, $(0,-2)$ and a scalar $H$ with U(1) charges
$(1,-1)$.
Then the Lagrangian is given by the gauge-covariant kinetic terms for all fields, plus the
$(\lambda_1\psi_1 H +\lambda_2 \psi_2 H^*)\psi$  Yukawa interactions.
The vacuum expectation value of $H$ breaks $G$ to the SM U(1)$_Y$,
leaving some particle massless (to be identified with the SM fermions and vectors,
that acquire a mass when the SM gauge group is broken)
and some other massive.
Imposing a $1\leftrightarrow2$ Z$_2$ permutations symmetry enforces $\lambda_1 = \lambda_2$ such that
the $H$eavy and $L$ight mass eigenstates
are $W_H=W_1-W_2$, $W_L = W_1 + W_2$,  $\psi_L = \psi_1 - \psi_2$, $\psi_H = \psi_1 + \psi_2$ and
the final Lagrangian has the form
${\cal L} \sim H^2 W_H^2 + H \psi \psi_H + W_H \psi_H \psi_L + W_L (\psi^2 + \psi_L^2 + \psi_H^2)$.
The lightest among the 
heavy vector $W_H$ or fermion $\psi_H$ is stable thanks to the Z$_2$ parity.
A more complicated construction can be applied to the non-abelian factors of the SM
allowing to get light chiral fermions, and again one
finds potential vector or fermion DM candidates.
UV completions the little-Higgs models usually render
the permutation symmetry anomalous~\cite{Hill}, ruining the stability of DM.
In the Universal Extra Dimension context, the problem is that truly universal extra dimensions
do not give chiral fermions, so that one needs to add extra structures such as boundaries:
divergent quantum corrections generate operators there localized, ruining the `universality'.}

In the light of these considerations, we can now list the possible polarization states that give rise to observationally
 nonequivalent spectra of
final SM particles, taking into account that DM annihilations cannot be observed
event-by-event, but only on a statistical basis, after propagation in the Galaxy.


\begin{itemize}
\item[1.] DM annihilations into two massive vectors, $W^+ W^-$ or $ZZ$. 
We focus on the DM DM singlet state $\mathscr{D}$, that exists for all possible DM spins.
It can couple to SM vectors in 3 possible ways, described by the following effective Lagrangians:
\begin{itemize}
\item[1a)] The effective interactions
$\mathscr{D} F_{\mu\nu}\epsilon_{\mu\nu\rho\sigma} F_{\rho\sigma}$ and
$\mathscr{D} F_{\mu\nu}^2$ both give rise to annihilation into vectors with transverse polarizations
(helicity correlations are different in the two cases, but cannot be observed).
Transversely polarized massive vectors decay in two light fermions with energy $E=x \, M$ as:
\beq dN/d\cos\theta = 3(1+\cos^2\theta)/8 \qquad \hbox{i.e.}\qquad
dN/dx = 3(1-2x+3x^2)/2,
 \eeq
where distributions are normalized to $N=1$,
$\theta$ is the angle between the momenta of the vector and of the secondary fermion
and  the last expression holds for $M\gg M_W$.



\item[1b)] The possible effective interaction
$\mathscr{D} A_\mu^2$ gives vectors with longitudinal polarizations.
Within an $\SU(2)_L$ invariant theory, such effective couplings arises when DM couples
to the components of the Higgs doublet that get eaten by the $W,Z$ vectors.
A longitudinal vector decays as 
\beq dN/d\cos\theta = 3(1-\cos^2\theta)/4 \qquad\hbox{i.e.}\qquad
 dN/dx = 6x(1-x).\eeq


\end{itemize}
Fig.\fig{mudecay}a shows the energy spectra of secondary fermions, in the limit $M\gg M_W$.

\begin{figure}[t]
\begin{center}
$$\includegraphics[width=0.45\textwidth]{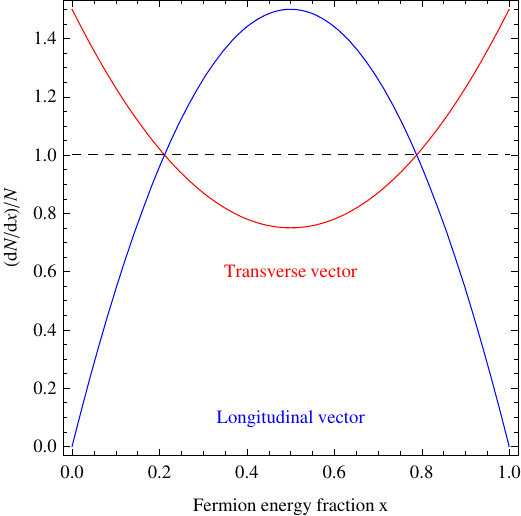}\qquad
\includegraphics[width=0.45\textwidth]{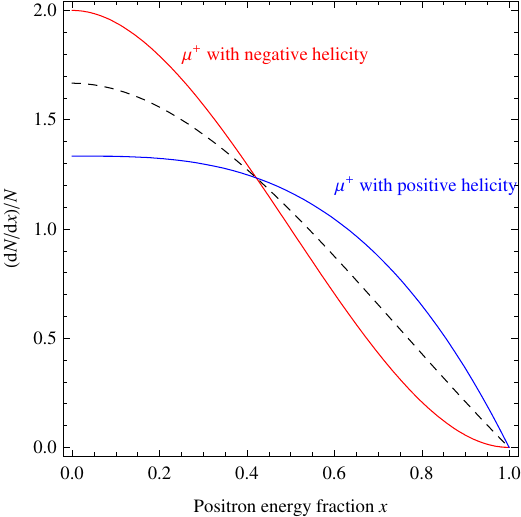}$$
\caption{\em Energy spectra of fermions $f$ produced by decays of a heavy 
vector (e.g.\ $f=e$ from $W\to e^+\nu_e$)
or of a heavy fermion (e.g.\ $f=e$ from $\mu^+ \to e^+\nu_e \bar\nu_\mu$)
 from polarized vector and fermion decays, in the limit $M\gg M_{W,Z}\gg m_{f}$.
 The dashed lines show the result for the unpolarized case.
\label{fig:mudecay}}
\end{center}\end{figure}

\item[2.] DM annihilations into the Higgs bosons.
We can again focus on $\mathscr{D}$, so that the effective interaction $\mathscr{D}h^2$ 
gives DM annihilations into two Higgs bosons.
Since they have no spin, there are no polarization issues.
Precision ElectroWeak data suggest a light Higgs boson mass: we here assume
that the Higgs mass is $115\GeV$ so that the Higgs dominantly decays into $b$ and $\tau$.
(Higgs decay modes into heavy vectors would be instead dominant if  $m_h> 2M_W$).
DM annihilations into $Zh$ will not be considered, as they are
are essentially given by the average of the $Z_L Z_L$ and $hh$ channels.



\item[3.] DM annihilations to SM fermions, DM DM  $\to f\bar{f}$, with mass $m_f$.
\begin{itemize}
\item[3a)] The DM two-body singlet state $\mathscr{D}$ can only couple as
$\mathscr{D} f_L f_R+\hbox{h.c.} = \mathscr{D} \bar\Psi_f\Psi_f$,
where Dirac and Weyl notations for fermions are related by $\Psi_f =(f_L,\bar f_R)$.
If $m_f$ is not negligible, the operator can have a complex coefficient with physical meaning;
the imaginary part gives rise to the $\mathscr{D}\bar\Psi_f\gamma_5 \Psi$ operator.
Anyhow, the h.c.\ implies that on average the produced fermions have zero helicity.
Furthermore, the Weyl notation makes clear that, unless DM is coupled to both $f_L$ and $f_R$,
the coefficient of this operator is suppressed by $m_f/M$.



\item[3b)] We next need to consider the vector two-body DM state, $\mathscr{D}_\mu$,
present if DM is a vector or a Dirac fermion.
It can couple to fermions in two different ways:
$\mathscr{D}_\mu[\bar f_L \gamma_\mu f_L]=\mathscr{D}_\mu[\bar \Psi_f \gamma_\mu P_L \Psi]$ 
or as
$\mathscr{D}_\mu[\bar f_R \gamma_\mu f_R]=\mathscr{D}_\mu[\bar \Psi_f \gamma_\mu P_R \Psi]$.
The first (second) operator means that $f$ fermions have negative (positive) helicity.
To see how helicity affects the energy distribution of final-state fermions,
let us consider weak decays $f\to f' f''{f}'''$ such as
$\mu^+\to \bar\nu_\mu e^+ \nu_e$ in the limit of $m_{f'},m_{f''},m_{f'''}\ll  m_f\ll M_W$.
In view of Fierz identities, $f'$ and $f''$ have the same energy distribution:
\beq dN/dx|_L=2(1-x)^2(1+2x),\qquad
 dN/dx|_R= 4(1-x^3)/3.\eeq


\end{itemize}
We can consider the two limiting cases of  $L$ and $R$ spectra because,
as plotted in Fig.\fig{mudecay}b, their average is the unpolarized spectrum of point 3a).

\end{itemize}

\begin{figure}[p]
\vspace{-1cm}
\begin{center}
$$\includegraphics[width=\textwidth]{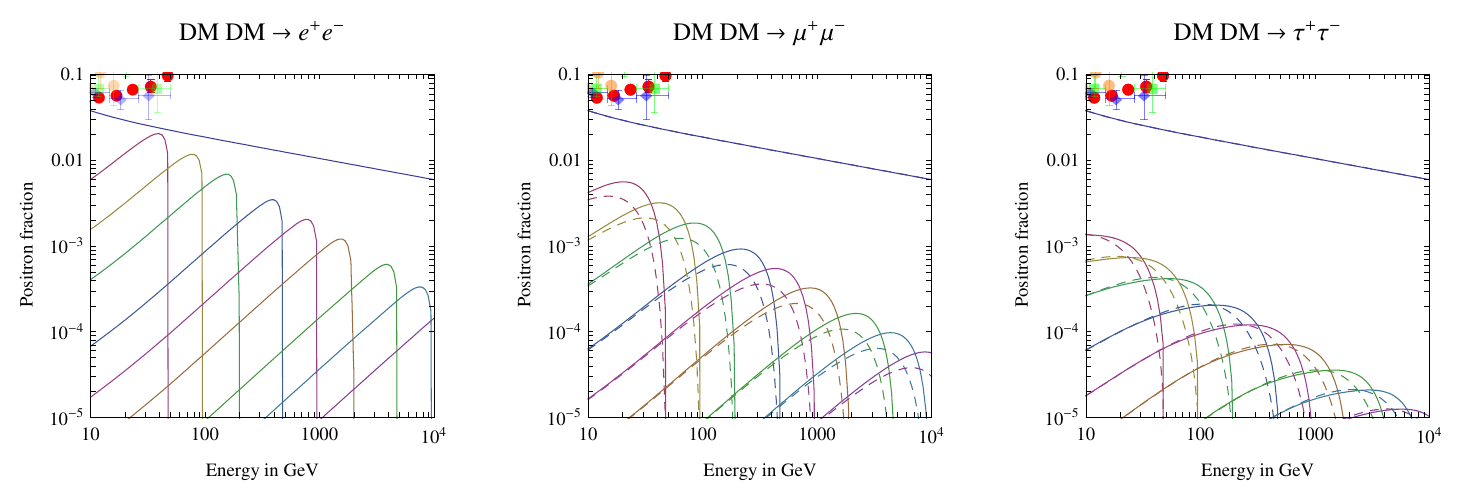}$$
$$\includegraphics[width=\textwidth]{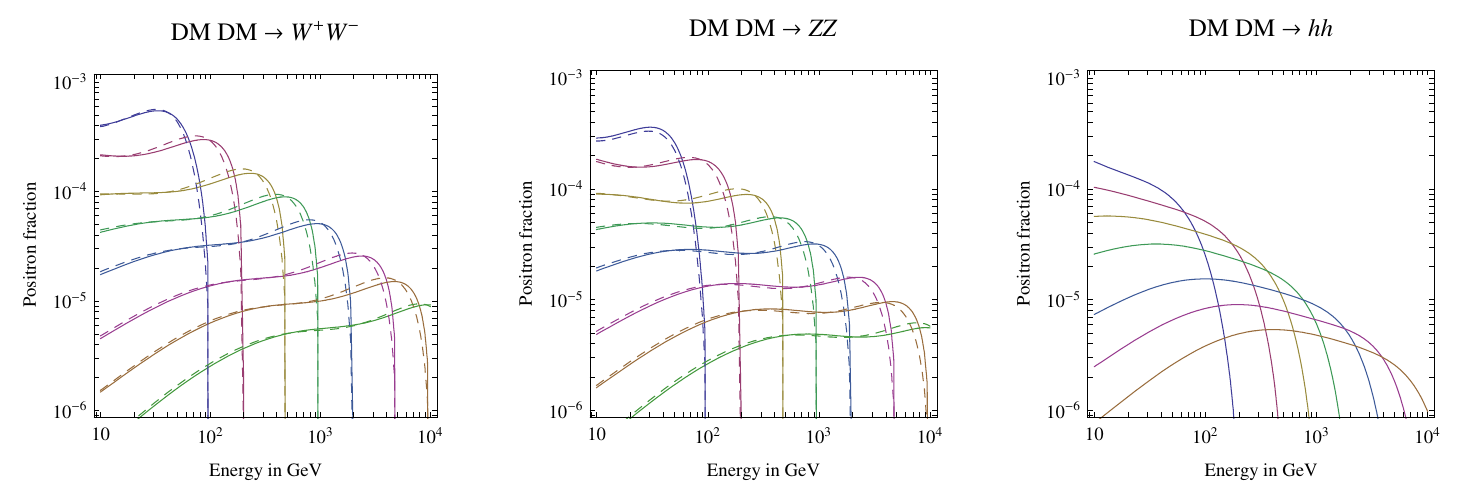}$$
$$\includegraphics[width=\textwidth]{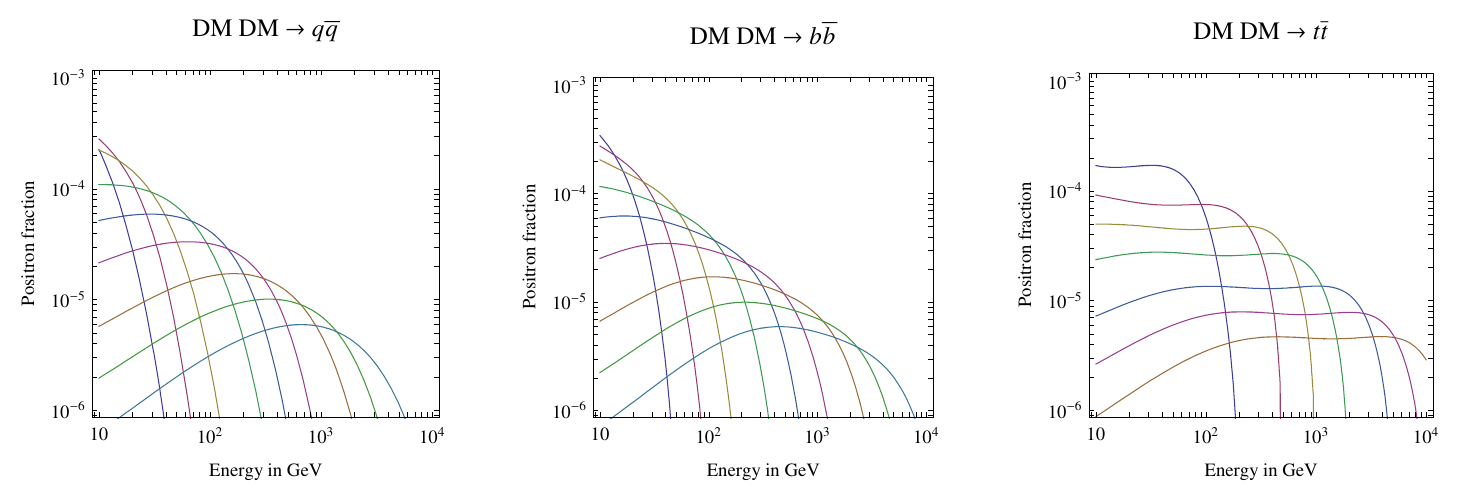}$$
\caption{\em Energy spectra of the positron fraction
$e^+/(e^-+e^+)$, as produced by DM annihilations into the various possible channels.
The DM masses are $\{50,100,\-200,\-500,\-1000,2000,5000,10000,20000\}\GeV$
and can be inferred from the energies at which the spectra drop to zero.
In this figure we take a NFW halo, the MED propagation model,
boost factor $B_e=1$ and the cross section $\sigma v=3\cdot 10^{-26}\cm^3/{\rm sec}$ na\"{\i}vely
suggested by cosmology.
For illustration, the upper plots show the expected astrophysical background (upper line) and the data, that can 
be fitted increasing $B_e\cdot\sigma v$.
The dashed curves in the $\mu,\tau,W,Z$ plots represent $\mu_R,\tau_R, W_L,Z_L$ polarizations,
while the continuous lines represent $\mu_L,\tau_L,W_T,Z_T$ polarizations.
\label{fig:espectra}}

\end{center}\end{figure}

\begin{figure}[t]
\begin{center}
$$\includegraphics[width=\textwidth]{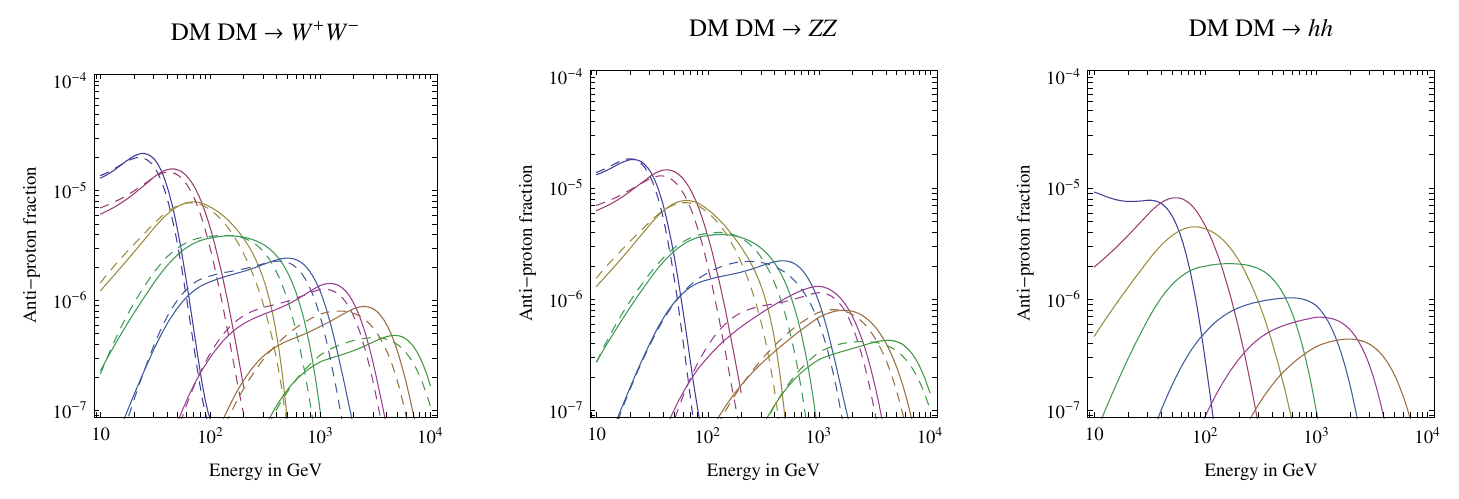}$$
$$\includegraphics[width=\textwidth]{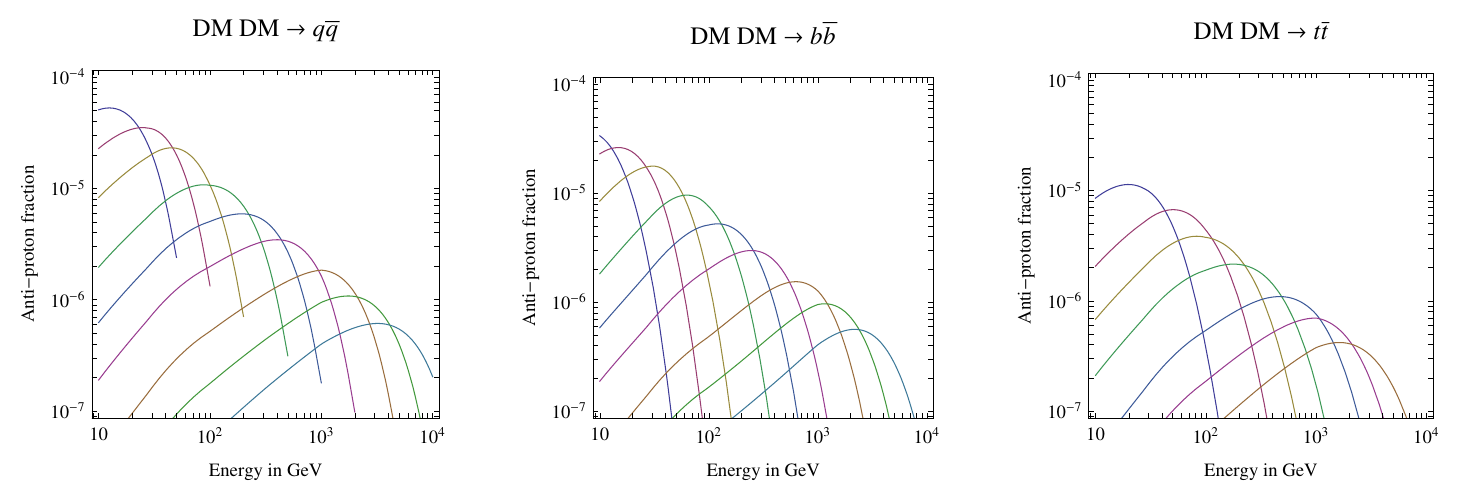}$$
\caption{\em The same as in Fig.\fig{espectra}  for the energy spectrum of the $\bar{p}/p$ ratio.
\label{fig:pspectra}}
\end{center}\end{figure}

Having precisely identified the primary two-body  states that can be produced in DM annihilations,
we next need to study how the spectra of final-state $e,p,\nu$ 
can be computed using MonteCarlo codes such as {\sc Pythia}~\cite{Sjostrand:2007gs},
that mostly neglect polarizations.
We computed polarized $W,Z$ decays using our own MC routine,
and decayed the resulting fermions using~{\sc Pythia}8.
For the $\mu$ spectra, we used analytic results.
For the $\tau$ spectra we used {\sc Tauola}~\cite{Was:2004dg}, 
interfaced with {\sc Pythia}6.
It would be interesting to repeat such computations with other MC tools
that employ different hadronization models, in order to get an estimate of the theoretical uncertainty.

In the simplest case of DM annihilations into $e^+e^-$,
we have taken into account radiation so that electrons and positrons have a smooth peak at $x\circa{<}1$
rather than a line at $x=1$, and the photon spectrum is
described by the Altarelli-Parisi splitting function
\beq \frac{dN_\gamma}{dx} = \frac{\alpha}{2\pi}\left[-1+\ln \bigg(\frac{4M^2}{m_e^2}(1-x)\bigg)\right]\frac{1+(1-x)^2}{x}.\eeq
These are the most interesting annihilation channels that will emerge from our later analysis,
and, especially in the $\mu$ case, polarizations are relevant.
For quarks, the polarization issue should be irrelevant compared to hadronization.
A possible exception is the $t$ quark.

In this way we obtained the $e$ and $p$ spectra at DM annihilations, 
plotted in Fig.\fig{espectra} and Fig.\fig{pspectra}.
These spectra will undergo deformations due to the propagation effects of the
charged particles in the galactic halo, to be discussed next.

\section{Propagation in the Galaxy and the astrophysical background}
\label{sec:propagation}
With the computed primary spectra in hand, we next need to consider DM annihilations in the Galaxy, and propagation of the resulting $e^\pm$ and $\bar p$ up to the solar system.
This is done proceeding as in~\cite{Delahaye:2007fr,Donato:2003xg,MDM3}, 
so we just briefly summarize the main points.
Three different halo models are considered: isothermal, NFW and Moore.
For each one of them, three different sets of parameters for propagation (different for $e^\pm$ and $\bar p$) are considered: MIN, MED and MAX.
The main physical parameter that discriminates between these configurations is the
thickness of the cylinder, centered on the galactic plane, inside which charged particles
such as $e^\pm$ and $\bar p$ diffuse.
The remaining parameters, such as the diffusion coefficients for $e^\pm,\bar{p}$, the energy losses for $e^\pm$ and the galactic convective wind speed for $\bar p$,
are fixed in order to reproduce the measured spectra of Cosmic Ray nuclei.

In line of principle, the astrophysical background spectra should be computed together with Cosmic Ray nuclei,
such that uncertainties are properly correlated.
We instead assume that the $e^+,e^-,\bar{p}$ background spectra can be freely renormalized,
and have independent $\pm0.05$ uncertainties in their energy slope
(namely, the central values used in~\cite{MDM3} are multiplied by $A \, E^p$ and the resulting $\chi^2$ is
minimized with respect to $p=0\pm 0.05$ and to $A$).
This mimics the main uncertainties in astrophysical backgrounds\footnote{We checked that this procedure reproduces reasonably well the uncertainty bands reported by more detailed analysis, see e.g. Fig.7  of \cite{BringmannSalati} for the case of antiprotons and Fig.4 of \cite{backgroundpositrons} for positrons. 
It seems to us unlikely that uncertainties in the electron astrophysical background (PAMELA is measuring it more accurately than ATIC~\cite{ATIC-2}) can produce the rising positron fraction
apparent in the PAMELA data.
}, produced by the Fermi mechanism
of acceleration, that typically generates power-law spectra (up to some cut-off) but does not predict its coefficient.
The positron fraction below 20 GeV has been measured very precisely, but is strongly affected by solar modulation,
which is not understood  well enough to allow a useful theoretical computation.
Therefore we assume that each data-point for the positron fraction has an independent
uncertainty given by the spread between different solar modulations, as plotted in~\cite{clem}.
This amounts to a $\pm6\%$ uncertainty at 10 GeV and $\pm 30\%$ at 1 GeV, and in practice means
that low energy data do not provide useful constraints on Dark Matter annihilations.

\begin{figure}[t]
\begin{center}
$$\includegraphics[width=\textwidth]{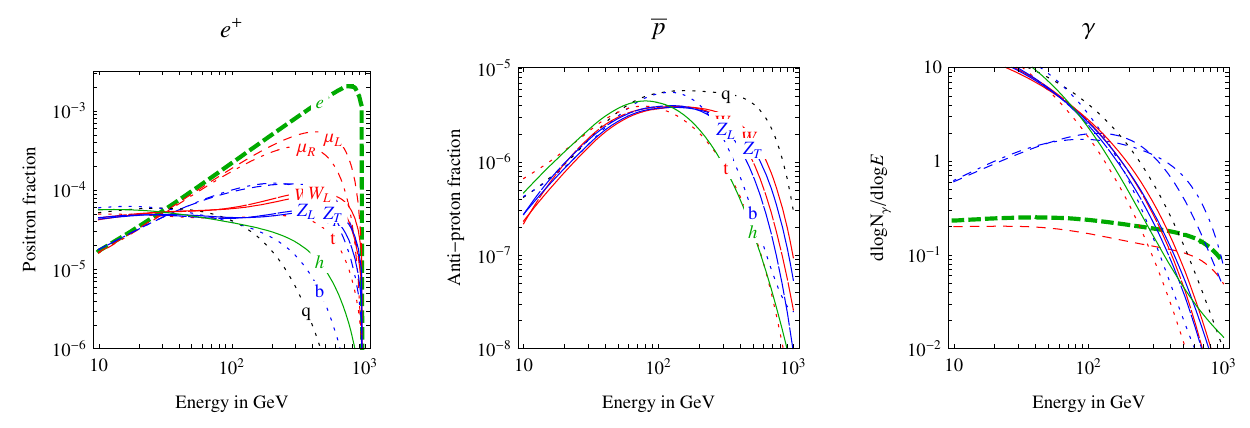}$$
\caption{\em Comparison of the electron (left) and proton (center) fractions and photon (right) fluxes produced by possible
DM annihilation channels, for $M=1\TeV$.
\label{fig:compspectra}}
\end{center}\end{figure}

In performing the fit, we smoothly scan over the intermediate propagation configurations and halo models,
within the boundaries described above. The MED configuration is sometimes considered as favored, but we do not attach a statistical meaning to this sentence.

Marginalizations over nuisance parameters and other statistical operations
are performed as described in Appendix B of~\cite{nureview}.
We will show plots of the $\chi^2$ as a function of the DM mass:  an interval at  $n$ standard deviations
corresponds (in Gaussian approximation) to $\chi^2 < \chi^2_{\rm min} + n^2$, irrespectively of
the number of data points.
We will not report the value of $\chi^2/{\rm dof}$ as it is a poor statistical indicator;
furthermore the number of dof is not a well-defined quantity when (as in the present case)
data-points with accuracies much smaller than  astrophysical uncertainties are effectively irrelevant.

\begin{figure}[t]
\begin{center}
$$
\includegraphics[width=0.65\textwidth]{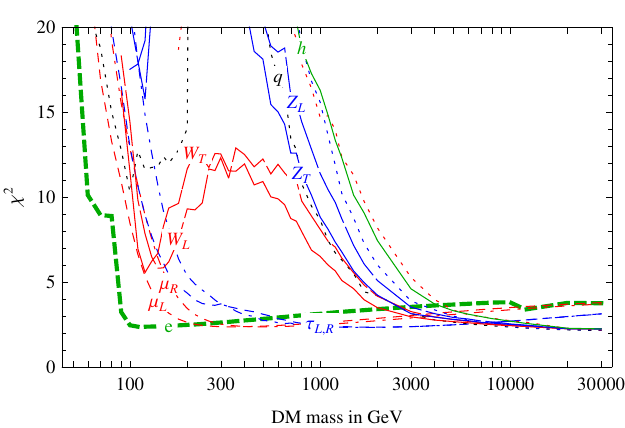}\qquad
\raisebox{8mm}{\includegraphics[height=6cm]{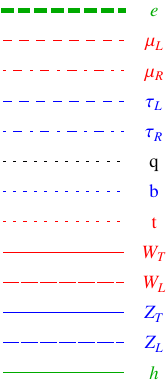}}$$
\caption{\em A fit of different DM annihilation channels to the 
 PAMELA positron fraction data only.
The labels on each curve indicate the primary annihilation channel.
\label{fig:fite}}
\end{center}
\end{figure}

\begin{figure}[t]
\begin{center}
$$\includegraphics[width=0.65\textwidth]{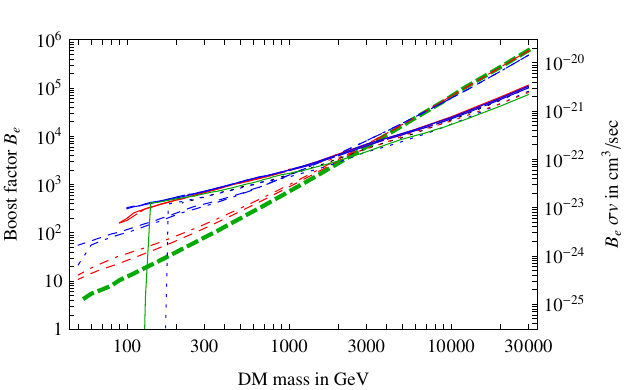}\qquad
\raisebox{8mm}{\includegraphics[height=5cm]{figs/legenda.pdf}}$$
\caption{\em Values of $B_e\cdot \sigma v$ (right axis) and of the boost factor $B_e$ (left axis, for
the standard $\sigma v = 3~10^{-26}\cm^3/{\rm sec}$) suggested by the PAMELA excess.
\label{fig:fitBooste}}\end{center}
\end{figure}

\section{PAMELA positron data}
\label{sec:PAMELApositrons}
We start our data analysis considering only the PAMELA $e^+/(e^++e^-$) observations (16 data points)~\cite{PAMELA}. 

Taking into account the DM distribution and positron propagation effects in the Galaxy, 
the energy spectra of the positron fraction
originating from different DM annihilation channels is plotted in the left panel of
Fig.~\ref{fig:compspectra} for the DM mass $M=1$~TeV. As expected, the 
most energetic positrons come from the pure leptonic channels and the 
softest spectra are produced in quark annihilation channels.

Fitting data as described in the previous section,
Fig.\fig{fite} shows how well the possible DM annihilations into two SM particles can fit the
PAMELA positron excess.
Fig.\fig{fitBooste} shows the boost factor $B_e$ (with respect to the cross section suggested by cosmology,
$\sigma v=3~10^{-26}\cm^3/{\rm sec}$) and $B_e\cdot \sigma v$  that best fits the PAMELA excess.
We see that DM annihilations into $e,\mu,\tau ,W$ can reasonably well reproduce the data for any DM mass,
while annihilations into $Z,t,q,b,h$ give a good fit for DM heavier than about 1 TeV.
It is perhaps interesting to note that, contrary to what commonly thought, the spectrum from $W^+W^-$ annihilations is not too flat to give a good fit of the quite steep PAMELA rise. At small masses (see e.g. the upper-left panel of Fig.~\ref{fig:samples}) a MIN configuration of the propagation parameters (and a proper variation of the background curve within the limits considered above) allows to fit the data. At large DM masses (see e.g. the lower-left panel of Fig.~\ref{fig:samples}) the low-$x$ portion of the primary spectrum is steep enough to do the job (as usual, $x=E/M_{\rm DM}$).


It is interesting to study if the PAMELA excess can be produced by supersymmetric DM.
Already at this stage one can see that it is necessary to give up some usual assumptions,
either the naturalness of supersymmetry (invoking multi-TeV DM) 
and/or DM astrophysics (invoking very large boost factors,
as in~\cite{Bergstrom:2008gr}) and/or that the DM is a thermal relic
(invoking some non-thermal DM cosmological production mechanism).
In this last case one interesting possibility emerges~\cite{Moroi:1999zb}. 
A pure Wino triplet 
annihilates predominantly into transverse $W^+W^-$ with
 $$\sigma v= \frac{g_2^4(1-M_W^2/M^2)^{3/2}}{2\pi M^2(2-M_W^2/M^2)^2},$$
that for $M\approx 100\GeV$ is precisely the value suggested by
the PAMELA $e^+$ excess, that would therefore be naturally produced with the boost factor
of order unity.

We now consider the PAMELA $\bar p$ data, showing that this latter and many other possibilities
are excluded because they lead to  large unseen $\bar p$ excess, as already illustrated in Fig.\fig{samples}.

\begin{figure}[t]\begin{center}
$$\includegraphics[width=0.65\textwidth]{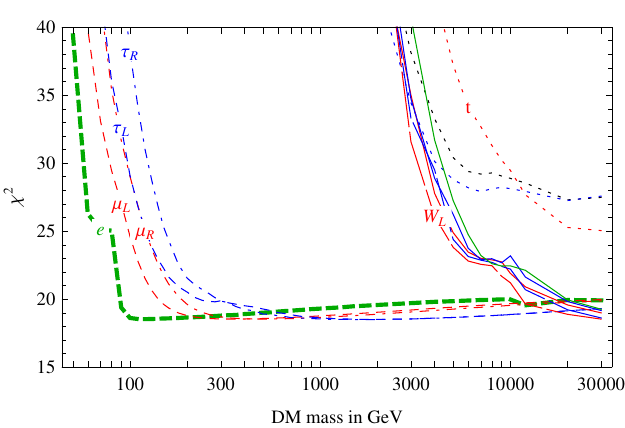}\qquad
\raisebox{6mm}{\includegraphics[height=6cm]{figs/legenda.pdf}}$$
\caption{\em Combined fit of different DM annihilation channels to the 
 PAMELA positron and PAMELA anti-proton data,
assuming equal boost factors and propagation model.
\label{fig:fitep}}
\end{center}
\end{figure}

\section{PAMELA positron and anti-proton data}
\label{sec:PAMELAantiprotons}
We address now the implications of the PAMELA $\bar p/p$ data~\cite{PAMELApbar}
(17 data points).
The middle panel of Fig.\fig{compspectra} shows the $\bar p/p$ fractions 
produced by the various primary channels. One can see that there are two basic features: 
i) annihilations into leptons do not give rise to protons;
ii) all other channels give rather similar proton energy spectra.
Comparison with the positron spectra reveals that the proton energy spectrum is
softer than the positron one produced in the leptonic and 
gauge boson annihilation channels.

The results of the fits are shown\footnote{In order to study how the fit of PAMELA data changes adding $\bar p/p$ data we assume equal boost factors and propagation models for positrons and antiprotons, as discussed in Sec.~\ref{sec:propagation}. The second assumption should be a good approximation, while the first one can easily fail when boost factors are very large, allowing to relax all our results.} in Fig.\fig{fitep}.

Consistently with what anticipated, since no excess seems present in the $\bar p/p$ ratio,
annihilation into leptons are not constrained as they do not produce antiprotons.
On the contrary, all other annihilations into quarks, vector and Higgs bosons
are significantly constrained, and allowed only if the DM particle is heavier than almost 10 TeV.
Only in such a case the proton excess lays at energies above those explored 
currently by PAMELA, while the low energy proton spectrum is consistent with the
background (see Fig.~\ref{fig:samples} for illustration).
The bound dominantly comes from high energy data points where the 
solar modulation is negligible.

The implications of the complementarity of  PAMELA  $e^+/(e^++e^-)$ and $\bar p/p$ data
on constraining new physics are evident. 
The light Wino considered in the previous section is excluded as a DM candidate
because its annihilations would induce a large unobserved antiproton excess.
Let us consider models where DM is a vector, the lightest Kaluza-Klein (KK) state 
or some other heavy replica of the photon or of the hypercharge boson or of the $Z$.  
They do  annihilate directly into leptons, but the direct annihilation rates into quarks
are of the same magnitude or larger, as quarks have 3 colors and similar charges.
For example the heavy KK of the hypercharge annihilates into fermions 
$f\bar{f}$ with $s$-wave cross section
\beq \sigma v = \frac{8\pi\alpha_Y^2}{9M^2}(Y_L^4+Y_R^4)N_c.\eeq
Such a  $\bar p$ excess cannot escape detection at PAMELA 
unless the DM particles have multi-TeV masses.
The same conclusion applies to the extra dimensional models in which the 
DM is the lightest KK state of neutrino as those annihilate predominantly to $WW$
final states. Similarly the DM candidates in the Little Higgs models must be very heavy
as they always have annihilation channels producing antiproton excess.
Therefore, the PAMELA data alone put very strong constraints on the mass of DM 
that have significant fraction of annihilations into quark or gauge boson channels.

\begin{figure}[t]\begin{center}
$$\includegraphics[width=0.65\textwidth]{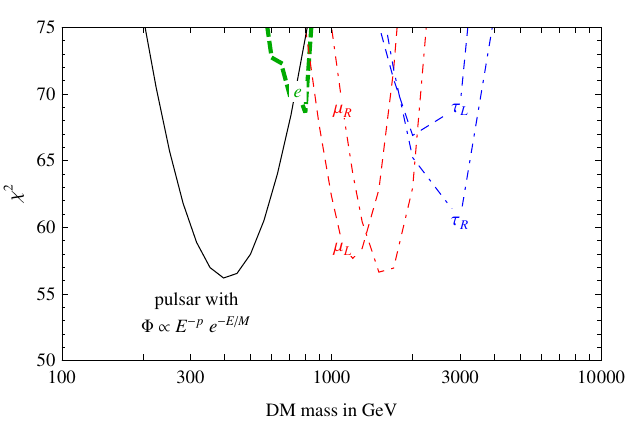}
\qquad
\raisebox{6mm}{\includegraphics[height=6cm]{figs/legenda.pdf}}$$
\caption{\em Combined fit of  PAMELA positron and balloon $e^++e^-$ data.
\label{fig:fitee}}
\end{center}
\end{figure}

\section{PAMELA positron and balloon $e^++e^-$ data}
\label{sec:balloons}
Various balloon experiments measured the spectrum of $e^++e^-$ cosmic rays
at energies higher than those where PAMELA can measure separately the $e^-$ and $e^+$ spectra.
If the $e^+/(e^++e^-)$ excess present in PAMELA data continues to rise in the same
way at higher energies, the positron fraction becomes of order unity at a few hundred GeV, giving rise to an excess in the $e^-+e^-$ spectrum. The most recent and precise measurements by ATIC-2~\cite{ATIC-2} and PPB-BETS~\cite{Torii:2008xu} indeed show a hint of an excess at such energies, which is not visible in the less precise EC data~\cite{EC}. The experimental situation is therefore quite open at these high energies, due also to the intrinsic difficulty of the measurements and the analysis. The ATIC-4~\cite{ATIC-4} experiment already collected new data, and GLAST~\cite{GLAST} is collecting new precise data, so that the possible structure in the  $e^++e^-$ spectrum should soon be clarified. Keeping this in mind, we explore the impact of the currently available results.
We fit ATIC-2, PPB-BETS and EC data, for a total of 37 data points.

As the DM annihilations necessarily should affect also the total $e^++e^-$ energy spectrum,
various issues deserve to be studied.
1)  Are  DM annihilations compatible with this hint?
2) Are DM annihilations compatible with this hint and with the PAMELA excess?
3)  Even if this $e^-+e^-$ hint  will not be confirmed, do $e^-+e^-$ data already
constrain DM annihilations?
The answer to all these questions is yes.

\medskip

A comparison of the $e^\pm$ energy spectra produced by the various annihilation channels
shown in Fig.\fig{compspectra}a indicates that only DM annihilations into $e,\mu$ or $\tau$
 can reproduce the $e^++e^-$ excess for a DM mass around 1 TeV.
In this case one obtains an excess in the positron fraction quantitatively similar
to the one present in the  PAMELA data.
In Fig.~\ref{fig:fitee} we present the combined fit to the lepton data. 
\begin{itemize}
\item DM annihilations into $\mu$ seem to give the optimal energy spectrum and the best fit,
as illustrated in the middle row of Fig.~\ref{fig:samples}.
Annihilations into $e$ ($\tau$) give a  slightly poorer fit, because of a too (not enough) steep spectrum.
The DM mass has to be around 1 TeV. 
Indeed, a heavier multi-TeV DM that annihilates into $e,\mu,\tau$ is excluded,
as a solution to the PAMELA excess, by $e^++e^-$ data.
A lighter DM that fits the PAMELA excess instead gives a $\circa{>}20\%$ drop in the $e^++e^-$ spectrum, not present in the ATIC-2 data: this is illustrated by the upper  row of Fig.~\ref{fig:samples}.

\item All other non-leptonic channels give an $e^++e^-$ excess which is too smooth to reproduce the hint in $e^++e^-$ data,
as illustrated in the bottom row of Fig.~\ref{fig:samples}: should the excess in the $e^++e^-$ spectrum not be confirmed, multi-TeV DM with dominant non-leptonic annihilations will be able to reasonably agree with a smooth $e^++e^-$
spectrum, while still producing the PAMELA $e^+/(e^++e^-)$ excess.

\end{itemize}
So, as the combined fit to all existing data prefers 1~TeV DM which annihilates only to leptons,
all the popular DM candidates discussed so far in this paper are disfavored.
If ATIC-4 will confirm the $e^++e^-$ anomaly, DM particles must have quite unexpected properties.

\section{Implications for DM direct detection}
\label{sec:direct}
We explore the connection between the cross section 
$\sigma_{\rm ann} v\sim 3\cdot 10^{-23}\cm^3/{\rm sec} \cdot (M/\TeV)^2$ 
suggested by the PAMELA indirect signal
and the cross section $\sigma_{\rm dir}$ probed by direct detection experiments.
Rather than studying precise but model-dependent implications, we 
try to delineate the qualitative generic pattern.
Depending on the DM annihilation channel, there are three different possibilities.
Taking into account a possibly large boost factor $B_e$ and/or Sommerfeld enhancement $S$, we estimate:
\begin{itemize}
\item[i)] DM annihilation into particles contained in nuclei, i.e.\ into quarks.
The qualitative connection between the cross sections is
\beq \sigma_{\rm dir} \sim \frac{m_N^2}{M^2}\frac{ \sigma_{\rm ann} v}{B_e S}\sim \frac{10^{-39}\cm^2}{B_e S}
\label{eq:dir1}
\eeq
although specific couplings can give one extra $m_N^2/M^2$ suppression.
The experimental bound on the spin-independent direct cross section is at the $10^{-41}\cm^2$ 
level~\cite{Ahmed:2008eu,Angle:2007uj}.
In the light of large boost factors $B_e$ and/or Sommerfeld correction $S$ required by PAMELA,
 eq.\eq{dir1} can be compatible with the present bounds.

\item[ii)] DM annihilations into  unstable SM 
particles not contained in normal matter, such as $W,Z,h,\mu,\tau$.
Then a direct cross section is generated at loop level, so that one expects a $\sigma_{\rm dir}$ smaller than in
eq.\eq{dir1} by one or two loop factors $\sim (4\pi)^{-2}$. Furthermore, the mass scale at the denominator
can be $M_W^2$ rather than $M^2$. The connection is therefore model dependent:
e.g.\ the Minimal Dark Matter quintuplet predicts $\sigma_{\rm dir}\sim10^{-44}\cm^2$~\cite{MDM1}.
 
\item[iii)] DM annihilations into electrons.  
 In order to observe the DM elastic scattering on electrons bounded in 
 atoms, DM $e^- \to$ DM $e^-$, one needs collisions with the most bound atomic electrons,
 that have quasi-relativistic momentum, $p\sim m_e$,
 such that the recoil energy $\Delta E \sim v p$ is above threshold~\cite{Bernabei:2007gr}.
The qualitative connection is
\beq \sigma_{\rm dir} \sim \frac{m_e^2}{M^2}\frac{ \sigma_{\rm ann} v}{B_e S}\sim \frac{10^{-45}\cm^2}{B_e S}.
\label{eq:dir3}
\eeq
The experimental apparatus must be
  able to detect electrons and must have low energy threshold for the electron
  detection. Such an experiment is, for example, DAMA~\cite{Bernabei:2008yi}.
 In order to explain the DAMA anomaly, the DM $e^-$ cross section should be
 $\sigma_{\rm dir}\sim 10^{-35}\cm^2(M/\TeV)$. Thus, in this scenario, the PAMELA data points to 
 very small direct cross section compared to the sensitivities of the present experiments.
\end{itemize}
Therefore DM motivated by the PAMELA anomaly is easily compatible with all the 
constraints from direct detection experiments.

\section{Implications for collider experiments}
\label{sec:collider}
We finally briefly discuss the implications for collider searches of the DM possibilities that our analysis has individuated as favored by the current data.
 
DM masses of $\cal{O}$(10) TeV are of course out of the reach of the LHC. 
DM particles with $\sim$1 TeV mass and weak interactions are at the limit of what can be detected by LHC experiments.  DM particles with $\sim$1 TeV mass which only couple to leptons would escape detection at LHC~\cite{Ball:2007zza}, unless accompanied by extra slightly heavier colored particles.
In the latter scenario the experimental signature is the usual missing transverse 
momentum carried away by DM in the cascade decays of the new model-dependent
colored particles.
Such a DM cannot be produced in the ILC either,  because of kinematical limitation on the center of mass energy $\sqrt{s}=0.5-1$~TeV. At future hypothetical $e^+e^-$ colliders like CERN CLIC or in the  $\mu^+\mu^-$ colliders with  $\sqrt{s}>2$~TeV, the direct production of DM particles is feasible. In particular, the PAMELA results on the annihilation cross section  DM DM$\to \ell^+ \ell^-$ imply a {\it lower bound} on the DM pair production cross
 section $\sigma(\ell^+ \ell^-\to \hbox{DM DM})$  in those colliders. Even if the required large boost factor
 $B$ is of astrophysical or cosmological origin, the production cross section of order 1~pb is large enough to observe such a DM at lepton colliders.

\section{Discussion and conclusions}
\label{sec:conclusions}

We studied if DM annihilations into two-body SM states
can reproduce the features that seem present in the
energy spectra of $e^+,e^-$, $\bar p$ cosmic rays recently 
measured at energies between about 10 GeV and 3 TeV by the PAMELA~\cite{PAMELA}, ATIC-2~\cite{ATIC-2} and PPB-BETS~\cite{Torii:2008xu} experiments.

We found that:
\begin{itemize}
\item Considering the  PAMELA positron data alone, 
the observed positron excess can be well fitted by DM annihilations into
$W,Z,e,\mu,\tau$ with any DM mass, 
and by DM annihilations into $q,b,t,h$ with a multi-TeV DM mass.

\item Adding the PAMELA anti-proton data suggests that the positron excess
can be fitted by DM annihilations into $W,Z,h$ (annihilations into $q,b,t$ give poorer fits)
only  with a multi-TeV DM mass,
unless boost factors or propagation strongly differentiate between $e^+$ and $p^-$.
DM annihilations into leptons are still viable for any DM mass.

\item Adding the balloon $e^++e^-$ data similarly suggests that DM annihilations into
$W,Z,h,b,q,t$ can fit the PAMELA excess only with a multi-TeV DM mass.
Annihilations into $e,\mu,\tau$ predict a feature in the $e^++e^-$ spectrum that,
for a DM mass  around one TeV, is compatible with the hint present in ATIC-2 and PPB-BETS data.
A TeV mass seems too heavy for explaining the DAMA anomaly~\cite{Bernabei:2008yi} compatibly with CDMS bounds in terms of inelastic DM scatterings~\cite{DAMAinel}.

\item  The annihilation cross section suggested by the
PAMELA data is a few orders of magnitude larger than what naturally suggested
by the cosmological abundance, unless DM formed sub-halos.
 This enhancement could be due to a Sommerfeld non-relativistic enhancement 
 of the annihilation cross section. Alternatively, this may indicate for non-thermal
 production of DM.

\item Present data start to be sensitive to the
polarization state of the DM annihilation products.

\end{itemize}
In the light of our results two solutions emerge for the DM.
\begin{enumerate}
\item The first possibility is a heavy DM, $M\circa{>}10\TeV$, that annihilates into $W^+ W^-$ or $hh$ (an example of a DM candidate of this sort is provided by the model of~\cite{MDM1}). 
This possibility is illustrated by the lower row in Fig.\fig{samples}, and will be tested by future $\bar p$ data.
\item
The second, more exciting, possibility is that the PAMELA excess has also been seen
in $e^++e^-$ data, by ATIC and PPB-BETS.
The forthcoming ATIC-4 and GLAST data should soon clarify this issue.
In such a case the best fit is obtained for $M\approx 1\TeV$ with DM annihilating into $\mu^+\mu^-$, 
and good fits are also obtained for $M\approx 800\GeV$ if DM annihilates into $e^+e^-$,
or $M\approx 2\TeV$ if DM annihilates into $\tau^+\tau^-$.
This is illustrated by the middle row in Fig.\fig{samples}.
The needed `boost times cross section' is $B_e \, \sigma v\sim 3~10^{-23}~\cm^2/{\rm sec}$.
\end{enumerate}
In both cases, a DM particle compatible with the PAMELA  anomaly 
has therefore unexpected properties. 
In general, $\bar p$ and $e^++e^-$ data seem to indicate that
DM masses around $100\GeV$ cannot fit the PAMELA anomaly, as
illustrated by the fit in the upper row of Fig.\fig{samples}.
The favorite DM candidate, the supersymmetric neutralino~\cite{Ellis:1983ew},
 cannot annihilate directly into light leptons with a large cross section~\cite{Bergstrom:2008gr}.
A  multi-TeV Wino is a possible supersymmetric candidate,
but supersymmetry would not naturally explain the breaking of the electroweak symmetry.
The vector or Dirac fermion DM candidates appearing in the models of 
extra dimensions~\cite{Servant:2002aq,Cheng:2002ej,Agashe:2004ci} 
and in the Little Higgs models~\cite{Cheng:2004yc} can annihilate directly into leptons.
However, at the same time they also annihilate  into quarks and/or 
into gauge and Higgs bosons. Therefore the masses of such DM candidates are also pushed to 
the multi-TeV region by the PAMELA antiproton data. 
Similarly, a LSP right-handed sneutrino with a sizable $\nu_R LH$ Yukawa coupling
annihilates into both  $L$ and $H$ (and consequently into longitudinal $W,Z$, that contain
the Goldstone components of the Higgs doublet $H$), unless lighter than $M_W$.
None of these theoretically motivated 
DM candidates mentioned here can explain the ATIC/PPB-BETS excess. 

A model of a thermal DM candidate that can explain both the PAMELA and the ATIC/PPB-BETS  excesses
with an order one boost factor can be constructed as follows.
We assume that lepton flavor $L_\mu-L_\tau$ is a spontaneously broken anomaly-free
U(1)$_L$ gauge symmetry,
with gauge coupling $\alpha_L \approx 1/50$ and vector mass $M_L\approx M_Z$.
The DM candidate is a Dirac fermion with mass $M\approx 1.5\TeV$, charge $q_L\approx 2$ under the extra U(1),
and no SM gauge interactions.
Its tree-level annihilation cross section relevant for cosmology,
$\sigma v = \pi (3 q_L^2 + q_L^4)\alpha_L^2/2M^2=3~10^{-26}\cm^3/{\rm sec}$ at $v\sim 0.2$,
is enhanced in astrophysics,
up to $\sigma v \circa{<}10^{-22}\cm^3/{\rm sec}$ at $v \sim 10^{-3}$,
by the Sommerfeld correction due to the extra U(1), as computed in Fig.~1a of~\cite{Sommerfeld2}.
The photon spectrum resulting from these DM annihilations into $\mu$ and $\tau$
is hard, as can be seen in the right panel of Fig.~\ref{fig:compspectra}.
The correction to the  magnetic moment of the muon,
$\delta a_\mu=+ g_L^2m_\mu^2/12\pi^2M_L^2$,
reproduces the observed discrepancy with respect to the SM, $\delta a_\mu=(3\pm1) 10^{-9}$~\cite{g-2}.
The new vector affects precision data as
$R_\ell\equiv \Gamma(Z\to \ell^+\ell^-)/\Gamma(Z\to e^+e^-) =1+ 3 \alpha_L f(M_L/M_Z)/4\pi$
with $f(0)=1$; this is compatible with the LEP measurements, $R_\mu=1.001\pm 0.003$
and $R_\tau=1.002\pm 0.003$~\cite{LEP}.
Furthermore  one loop-effects generate a kinetic mixing of the new vector with the photon,
$\theta = e\, g_L \ln (m_\tau/m_\mu)/6\pi^2 \approx 0.005$,
that gives rise to a cross section for direct detection,
$\sigma_{\rm SI}=4\pi q_L^2\alpha_L\alpha m_N^2 \theta^2/M_V^4\sim  10^{-42}\cm^2$, around the present bound.
Reducing $M_V$ by a factor of few leads to a too large $\sigma_{\rm SI}$,
and reducing $M_V$ below about 1 GeV also to unseen DM annihilations in first structures~\cite{first}.

\bigskip

Finally, a third, less exciting possibility is that the features that appear in the $e^+$ and $e^+ + e^-$ spectra are
not due to DM annihilations but to some astrophysical object, such as a pulsar.
Indeed it is expected that pulsars produce a power law spectrum 
(with index $p$ around 1.5--2) of mostly electron-positron pairs, with a cut-off.
The black line in  Fig.\fig{fitee} shows that a flux
$\Phi_{e^-} = \Phi_{e^+}=A\cdot  E^{-p} e^{-E/M}$ with $p\sim 2$ and 
cut-off at $M\sim 300\GeV$ does fit the $e^+,e^-$ data,
giving a peak  less sharp than what hinted to by ATIC-2 data.
It is remarkable that pulsars naturally give an excess only in $e^\pm$, with the observed slope $p$.
While the total energy and total power emitted by a pulsar can be
reliably computed assuming that they are respectively dominated by rotation and magnetic dipole radiation,
the factors $A$ and $M$ cannot be computed without a detailed model of the $e^\pm$ acceleration mechanism.
Furthermore, while the known nearby pulsars Geminga or B0656+14 are the main candidates,
unknown pulsars can contribute.
According to~\cite{pulsar}, young nearby pulsars tend to give a cut-off $M$ at energies higher than what suggested by
ATIC/PPB-BETS, while older pulsars tend to give a too low flux, as $e^\pm$ had time to diffuse away.
An old intense pulsar no longer emits $\gamma$, 
so that it is difficult to distinguish it from DM annihilating into leptons;
hopefully angular anisotropies of the $e^+,e^-$ fluxes will help~\cite{pulsar}.

\small

\paragraph{Acknowledgements} 
We thank Franco Cervelli,
Dario Grasso, Nicolao Fornengo and Vyacheslav S. Rychkov for very useful discussions.
This work was supported by the ESF Grant 6140 and by the Ministry of Education and 
Research of the Republic of Estonia.
A.S.\ thanks the EU project ENTApP working group 2 for financial support.
The work of M.C. is partially supported by the french ANR grants DARKPHYS and PHYS@COL\&COS.
We thank the EU Marie Curie Research \& Training network "UniverseNet" (MRTN-CT-2006-035863) for support.

\bigskip

\appendix

\footnotesize

\begin{multicols}{2}

\end{multicols}

\newpage\normalsize

\begin{figure}[t]
$$
\includegraphics[width=0.36\textwidth]{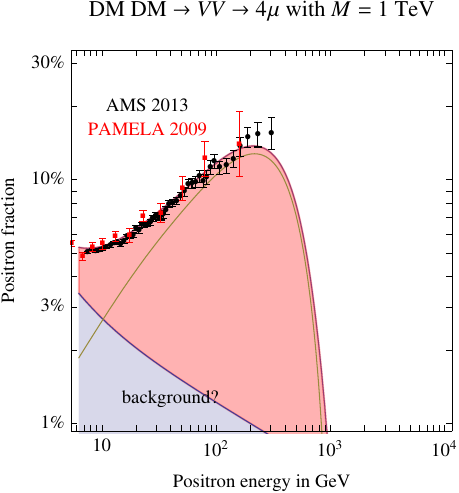}\qquad
\includegraphics[width=0.55\textwidth]{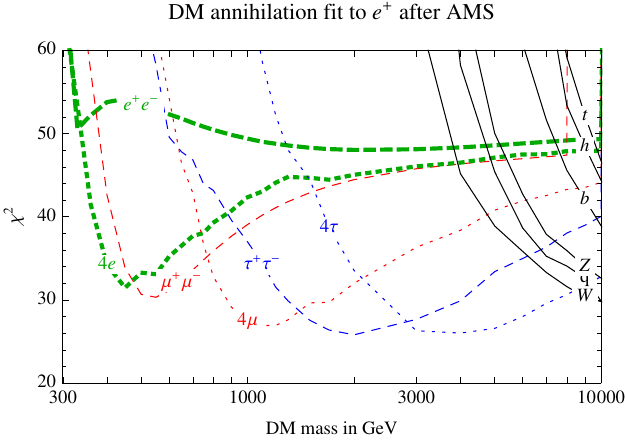}
$$
\caption{\em {\bf Left}: AMS data and  DM best fit.
{\bf Right}:
Fits of different DM annihilation channels to the 
 AMS and PAMELA positron fraction data only.
The labels on each curve indicate the primary annihilation channel.
\label{fig:fitAMSe}}
\end{figure}

\begin{figure}[t]
$$
\includegraphics[width=0.37\textwidth]{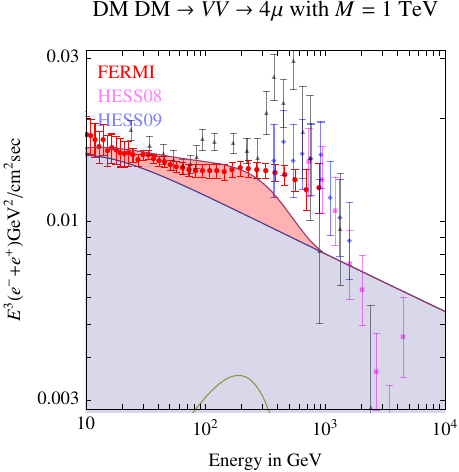}\qquad
\includegraphics[width=0.55\textwidth]{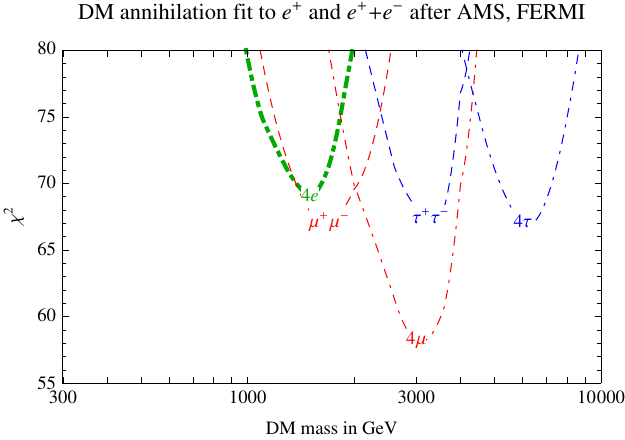}
$$
\caption{\em {\bf Left}: FERMI $e^++ e^-$ energy spectrum compared to the DM best fit
favoured by AMS $e^+$ data.
{\bf Right}:
Fits of different DM annihilation channels to the $e^+$ data and to the $e^+ + e^-$ data.
The labels on each curve indicate the primary annihilation channel.
\label{fig:fitFERMI}}
\end{figure}


\begin{figure}
$$
\includegraphics[width=0.48\textwidth]{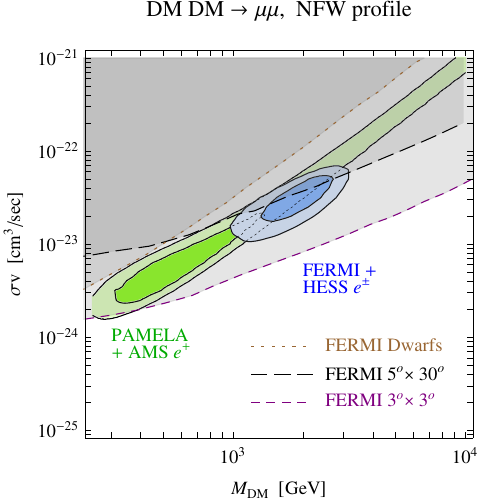}\qquad
\includegraphics[width=0.48\textwidth]{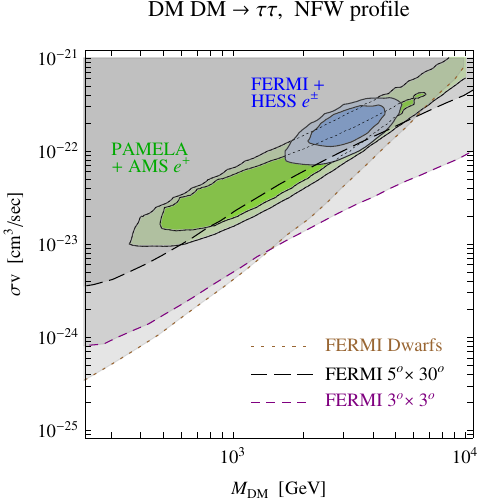}
$$
$$
\includegraphics[width=0.48\textwidth]{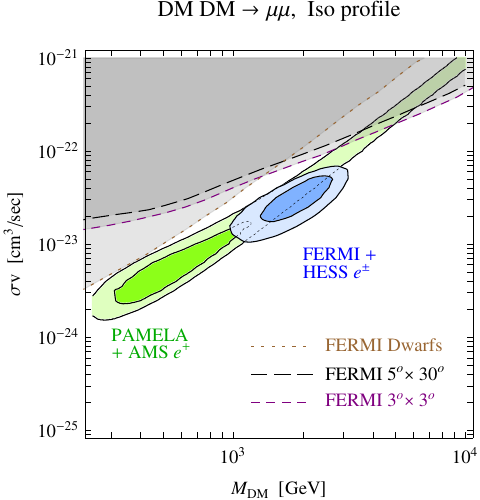}\qquad
\includegraphics[width=0.48\textwidth]{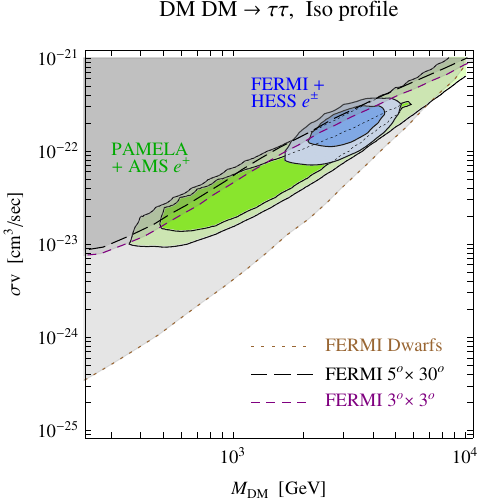}
$$
\caption{\em Regions of the ($M, \sigma v$) plane
favoured by AMS and PAMELA measurements of the positron fraction
(green) and by FERMI and HESS measurements of the electron and positron fluxes (blue), both at 3$\sigma$ and 5$\sigma$, for the $\mu^+\mu^-$ (left) and $\tau^+\tau^-$ annihilation channel   (right), compared to representative $\gamma$-ray constraints for the NFW (upper) and isothermal (lower) DM density profiles.
\label{fig:fitplane}}
\end{figure}

\section*{Addendum: AMS 2013 positron data}\label{AMSe}
The AMS collaboration confirmed the PAMELA measurement of a cosmic ray $e^+/(e^++e^-)$
 fraction rising with energy~\cite{AMS}.
AMS data are shown in fig.\fig{fitAMSe}a.
Consequently, our DM interpretation of the PAMELA excess remains valid (in particular the needed large DM boost factors).
In fig.\fig{fitAMSe}b we fit the AMS and PAMELA positron data  
to various DM annihilation channels.
The uncertainties on the DM profile density, on the $e^\pm$ propagation in the galaxy,
on the astrophysical background, etc, are taken into account as described in the main text.
The $4\ell$ channels added in fig.\fig{fitAMSe}b 
(with respect to our previous fig.\fig{fite}) 
correspond to DM annihilations into two speculative light vectors 
which decay into light leptons.
The significantly improved precision of the AMS data leads to just a few new notable implications:

\begin{itemize}
\item[i)]  Leptonic DM annihilations are now demanded by positron data only, without invoking $\bar p$ data.
Other channels are disfavoured, unless DM is heavier than a few TeV.
\end{itemize}
This is the case, for example, of the DM DM $\to W^+W^-$ annihilation channel
characteristic by wino DM, which predicts a $e^+$ rise with a spectrum different from what is now measured.
This channel was already found to be
strongly disfavoured in the previous version of this paper
from the lack of a $\bar p$ anomaly.

\begin{itemize}
\item[ii)] The hint of a flattening in the positron fraction suggested by AMS
now favours a DM mass below about 1 TeV with about $3\sigma$ statistical significance, depending on the DM annihilation channel.
Fig.\fig{fitAMSe}a shows the best fit.
\end{itemize}
DM models predict that the flattening of the $e^+/(e^++e^-)$ energy spectrum
should turn into a drop of the positron fraction, because 
annihilations of DM with mass $M$ cannot produce $e^\pm$ with energies larger than $M$.
This drop is not supported by FERMI data on the $e^+ + e^-$ spectrum,
which (within the $\approx 6\%$ uncertainties on the data-points) is smooth up to about 1 TeV, 
with no hint of a  $\approx 25\%$ decrease.
Fig.\fig{fitFERMI}a exemplifies how the drop below 1 TeV hinted by AMS $e^+$ data within DM models
would give a
decrease not present in the FERMI data.
On the contrary, the global set of $e^++e^-$ data hints to a decrease at a larger energy.
Fig.\fig{fitFERMI}b (which updates fig.\fig{fitee} in the original version of this paper)
shows that, as a consequence,
in a global fit, FERMI $e^+ +e^-$ data
favor a DM about 3 times heavier than AMS $e^+$ data.

Rather that indulging in speculations about astrophysical backgrounds that could partially compensate for the DM drop,
we point out that AMS can easily 
clarify the issue by performing a very precise measurement of the  $e^+ +  e^-$ spectrum.

\medskip

Finally, we comment on the compatibility of the DM interpretation of the rising positron fraction with constraints
from observations of $\gamma$ cosmic rays. In fig.\fig{fitFERMI} we show representative $\gamma$-ray bounds (the constraints are taken from~\cite{gamma,FERMI}, more recent analyses find similar or slightly more stringent bounds). We see that the new fit region of AMS shows some tension with $\gamma$-ray data (in the case of annihilations into
$\mu^+\mu^-$) or it is rather clearly excluded (in the case of annihilations into $\tau^+\tau^-$).
In the upper row we have chosen a benchmark NFW galactic Dark Matter profile; choosing the shallower isothermal
profile (lower row) the constraints become looser. 
%

Furthermore, we recall that observations of the cosmic microwave background (CMB)
imposes bounds on DM annihilations (based on the fact that they
would have re-ionised the primordial universe) that disfavor at various degrees and for most channels the DM interpretation of the positron excess~\cite{CMB}.

\footnotesize
\begin{multicols}{2}

\end{multicols}

\end{document}